\documentclass[aps,prl,twocolumn]{revtex4-2}
\usepackage{graphicx}
\usepackage{amsmath}
\usepackage{amsfonts}
\usepackage{amssymb}
\usepackage{xcolor}

\newcommand{\beq}{\begin{equation}}
\newcommand{\eeq}{\end{equation}}
\newcommand{\beqa}{\begin{eqnarray}}
\newcommand{\eeqa}{\end{eqnarray}}

\begin{document}

\title{Thresholdless stochastic particle heating by a single wave}

\author{F. Sattin$^{1}$, D.F. Escande$^{1,2}$}
\email{fabio.sattin@igi.cnr.it,dominique.escande@univ-amu.fr}
\address{
$^{1}$ Consorzio RFX (CNR, ENEA, INFN, Universit\`a di Padova, Acciaierie Venete SpA), Corso Stati Uniti 4, 35127 Padova, Italy \\
$^{2}$  Aix-Marseille Universit\'e, CNRS, PIIM, UMR 7345, Marseille, France}

\begin{abstract}
Stochastic heating is a well-known mechanism through which magnetized particles may be energized by low-frequency electromagnetic waves. In its simplest version, under spatially homogeneous conditions, it is known to be operative only above a threshold in the normalized wave amplitude, which may be a demanding requisite in actual scenarios, severely restricting its range of applicability. In this work we show, by numerical simulations supported by inspection of the particle Hamiltonian,  that allowing for even a very weak spatial inhomogeneity completely removes the threshold, trading the requirement upon the wave amplitude with a requisite upon the duration of the interaction between wave and particle. The thresholdless chaotic mechanism considered here is likely to be applicable to other inhomogeneous systems.
\end{abstract}

\maketitle

{\it Introduction.}
Stochastic heating by electrostatic or electromagnetic waves is a well-known mechanism often invoked as the cause of the energization of magnetized particles. By itself, it may be produced by waves whose frequencies are above or below the particle cyclotron frequency, by monochromatic or broad wave spectra. Actually, chaotic would be a more appropriate adjective than stochastic, since randomness is not a mandatory requisite. 
In the following we will consider the scenario of a single plane wave below the cyclotron frequency; this mechanism has been theoretically studied in several papers  \cite{drake81,chia96,johnson01,chen01,white02,lv07,bour08,vranjes09,Lu09,Sheng09,Ma09,voitenko10,chandran10,
wang11,Gao12,sun14,liang17,Yu17,yoon18,escande19,Yoon21},
 and compared with measurements coming from astrophysical, laboratory and high-temperature fusionistic plasmas \cite{mcchesney87,mcchesney91,sanders98,Chatto98,enge10,vranjes10,stasiewicz13,holst14,klein16,
 artemyev17,vech17,martinovic19,martinovic20}. 
Its paradigmatic version consists of a particle embedded in a homogeneous magnetic field, which interacts with a single monochromatic wave propagating perpendicularly to the field itself. Electrostatic waves, such as lower hybrid, drift waves, or electromagnetic waves such as whistlers and obliquely propagating Alfv\'en waves, are often considered. The model admits a 1-dimensional Hamiltonian formulation for a particle of charge $q$ and mass $m$, embedded in a magnetic field of intensity $B$ aligned along the direction $z$, which interacts with a plane wave of wavenumber $k$ and frequency $\omega$ propagating along $x$. In dimensionless form 
\beq
H =  {p^2 \over 2 } + {x^2 \over 2 } - A \sin ( x - \omega t)
\label{eq:1}
\eeq
In (\ref{eq:1}) lengths are normalized to $k$, time and frequencies to the cyclotron frequency $\Omega = q B/m$. The expression of the normalized wave amplitude $A$ depends on the kind of wave considered.  For an electrostatic wave, e.g., it writes $A = (k/\Omega)\times (F/B)$, where $F$ is the component of the wave electric field perpendicular to $B$. \\
When $\omega \ll 1$ one naively expects adiabatic behaviour, with the particle adapting to the instantaneous phase of the wave.  On the contrary, analysis of the system (\ref{eq:1}) has shown that, whatever the smallness of $\omega$, the particle experiences non-adiabatic behaviour with chaotic motion, non-conservation of the magnetic moment, and irreversible transfer of energy, provided that $A$ fulfils the threshold condition $A \ge 1$ \cite{escande19}. The rationale is that, above the threshold, a separatrix curve oscillating at frequency $\omega$ appears in the particle phase space. Particle orbits are shown to cross quasi-periodically these slowly pulsating separatrices. Since the orbit along the separatrix has an infinite period--effectively much longer than the pulsation one--each separatrix crossing cancels adiabatic invariance. This case is an instance of the {\it neo-adiabatic theory} developed in the eighties-nineties \cite{bruh89,elskens91,elskens93,lieberman92,bazzani99} (and also \cite{drake81,artemyev11,neishtadt19}). 
Below the threshold for the appearance of the separatrix, if one looks at the particle degrees of freedom, they may still suffer fluctuations apparently resembling heating in the presence of the wave train, but when the latter vanishes the particle energy lands back to its original value (see, e.g., Fig. 3 of \cite{escande19}). This kind of apparent transient heating is named sometimes ``pseudoheating'' in the literature \cite{wang09}.   \\
The chaotic mechanism has found widespread applications, yet the need for a finite threshold puts severe constraints to its validity. Just to cite two examples, where stochastic heating might be a potential player, but the expected wave amplitude appears too small, we mention electron heating at the Earth's magnetosphere \cite{stasiewicz20, pezzi22}, and ion heating during magnetic reconnection in toroidal laboratory devices \cite{ganga08,tangri08,ficksel09,magee11,kumar13,cartolano14,ren11}.

So far, Hamiltonian Eq. (\ref{eq:1}) has almost always been studied under the hypothesis of a perfectly homogeneous environment, which is unrealistic (Of course, the existence of any wave entails some sort of inhomogeneity. For the scope of this work, a homogeneous system is one where  {\it the parameters}  $A,k, \omega$  appearing in Eq. (\ref{eq:1}) remain constant. 
We are aware of few exceptions: Gell and Nakach \cite{gell80} and Albert \cite{albert00} studied the effect of a sheared magnetic field upon particle energization by waves; Ryabova and Shklyar addressed the effect of the longitudinal variation of the magnetic field \cite{ryabova83}. All cases were not in the low-frequency regime though. It was observed that accounting for some sort of inhomogeneity does lower the threshold for the occurrence of chaotic motion. \\
The purpose of this work is to show that, under more realistic conditions than present in Hamitonian (\ref{eq:1}) with $\omega =$ constant, the threshold condition on $ A$ is not actually required. Heating may take place at any value of the wave amplitude; the underlying physical mechanism is still the same: adiabaticity breakup is forced by the appearance of a separatrix in phase space, which is crossed by the particle. However, the separatrix appears gradually, thus adiabatic dynamics holds in the first stage of the particle evolution, and the crossover to non-adiabatic dynamics is gradual and may be relatively slow. 
This work shows by numerical analysis and analytical inspection of Eq. (\ref{eq:1}) that even a small spatial inhomogeneity in $\omega$, whose length scale is much larger than  the Larmor radius, changes radically the behaviour of the system, removes the threshold, and allows for a non-adiabatic behaviour to appear at any value of the wave amplitude. 

\begin{figure}[h]
\includegraphics[width=8.5cm]{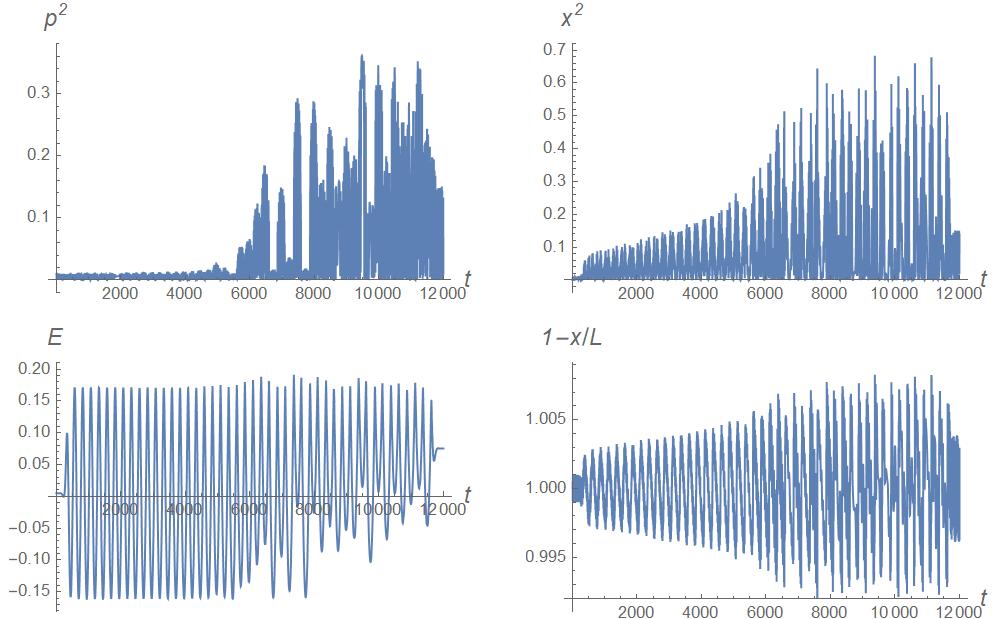}
\caption{Upper row, time traces of $x^2, p^2$. Lower row, left plot: time traces of the total energy $E = H$ (blue curve). The fast oscillations are due to the wave. The particle is initialized with a very small energy, hence oscillations are almost symmetrical with respect to zero energy. With time, the particle acquires a finite average amount of energy. Right plot, time trace of $ 1-x/L$.
Initial conditions are $x(0) = 0, p(0) = 0.1$, frequency $\omega^{(0)} = 1/40, L = 100, A = 1/6$. Temporal shaping is made with $\Delta = 20 \pi, t_s = 5\times \Delta, t_{max} = 12\times 10^3, t_e = t_{max} - 5 \times \Delta$.}
\label{fig:1}
\end{figure}

{\it Numerical Investigations.} 
We allow for a weak spatial dependence in the wave frequency $\omega$ appearing in (\ref{eq:1}), taken as linear for simplicity:  $\omega(x) = \omega^{(0)} (1 - x/L)$ with $|x/L| \ll 1$. For instance, were the case of an Alfv\'en wave, one could think of a plasma density with a mild variation along the x direction. 
We investigate the particle trajectories by integrating in time Hamilton equations of motion. For greater accuracy symplectic algorithms were employed, either the partitioned Runge-Kutta algorithm built into MATHEMATICA software or that described in \cite{atela92}.
In order to avoid spurious high-frequency forcings, we switch on and off the wave through a shape function: $A \to A f(t)$, where $f(t)$ is the double-sigmoid function $f(t) = 1/4 \left[ 1+ \tanh \left((t - t_s)/\Delta \right)\right] \times \left[ 1 + \tanh \left( (t_e - t)/\Delta \right) \right]$.  
Numerical integration runs between 0 and $t_{max} > t_e > t_s$. The interaction time is $t_e - t_s$ and $\Delta$ a typical rise time scale, being careful to  choose $\Delta \gg 2 \pi \Omega^{-1}$. 
The figures below show some samples of our results. Figure (\ref{fig:1}) plots time traces for the coordinates $x^2, p^2$, the total energy $E$ of Eq. \ref{eq:1}, and the inhomogeneity parameter $1-x(t)/L$  as well as $f(t)$. The parameters employed are given in the figure caption. 
We see that the particle remains throughout the whole simulation quite close to its initial position: $|x|/L$ steadily increases but remains $ \leq 0.5\%$; despite this there is a clear change in the particle motion visible after $ t = 5000 - 6000$ time units, with a net variation of its average energy.

\begin{figure}[h]
\includegraphics[width=8cm]{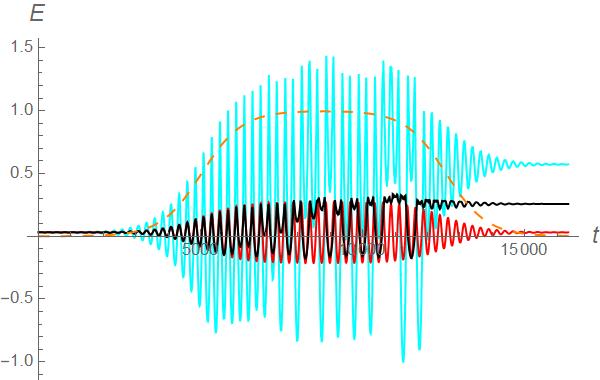}
\caption{Time traces of the energy $E$ for different conditions. Black curve, $L = 100, A = 1/4$; red curve, $L = 1000, A = 1/4$;  cyan curve, $L = 1000, A = 5/4$. In all cases $\omega^{(0)} = 1/40, x(0) = 0, p(0) = 1/4$.The dashed orange curve is the modulation function $f(t)$ employed, with $t_s = 20 \times T_{wave}, t_e = 50 \times T_{wave}, \Delta = 5 \times T_{wave}, T_{wave} = 2 \pi/\omega^{(0)}$. } 

\label{fig:2}
\end{figure}

The heating effect can be clearly assessed in Figure (\ref{fig:2}), which shows a few time traces of the particle energy $E$ with varying $A$ and $L$. We observe a net energy gain for sub-threshold amplitude and moderate inhomogeneity parameter $L$ (black curve). As expected, the energy gain diminishes as homogeneity is restored (red curve). If we raise the wave amplitude above the threshold (cyan curve), however, heating still occurs. 

\begin{figure}[h]
\includegraphics[width=8.5cm]{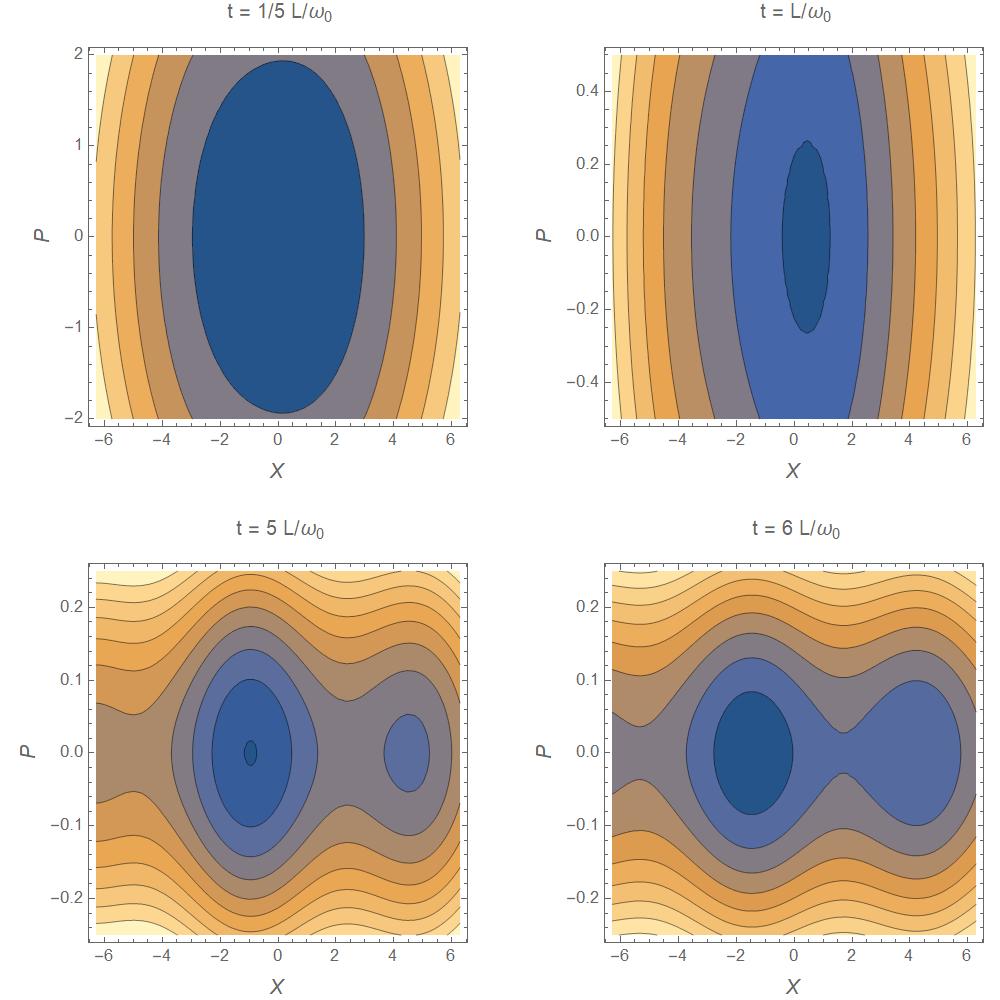}
\caption{Contour plot of the Hamiltonian $H'$ (Eq. \ref{eq:Hp}) at four different times, from $ \omega^{(0)} t / L \ll 1$ to $ \gg 1$. At the latter times, the separatrix is present. }
\label{fig:3}
\end{figure}

{\it Emergence of chaotic dynamics.} The previous numerical part showed that sub-threshold energization does occur. Here, we show why.    \\
First of all, we rewrite Eq. (\ref{eq:1}) with the help of the auxiliary parameter $q(t) = 1 + \omega^{(0)} t/L$:
\begin{equation}
H = {p^2 \over 2} + {x^2 \over 2} - A \sin \left( q(t) x - \omega^{(0)} t \right) 
\label{H2}
\end{equation}
Then we make a change to new canonical variables $(X,P)$ through the generating function
$F(X,p,t) = - {X p/q}$. One can see that $X = q x, P = p/q$ and the new Hamiltonian $K$ writes
\begin{equation}
K = { (q P)^2 \over 2 } + {X^2 \over 2 q^2} - A \sin \left[ X - \omega^{(0)} t \right] - {X P \over q} {dq \over dt}
\label{eq:Hp}
\end{equation}
This way, the $x$-dependence from the argument of the trigonometric term has been removed, making the expression of the Hamiltonian closer to that the original homogeneous problem. 
 Finally, introducing the new time $\tau$ through $d\tau = q^2 dt $ we arrive to the equations of motion
\begin{eqnarray}
d_\tau X &=& P + {X \over q^3} {dq \over dt} \nonumber  \\ 
d_\tau P &=& - {X \over q^4} - {P \over q^3} {dq \over dt} + {A \over q^2} \cos \left[X - \omega^{(0)} t(\tau) \right]
\label{eq:xp}
\end{eqnarray}
We remind that we are considering an initial-value problem, hence only $t \ge 0$ makes sense. When $ \omega^{(0)} t /L \ll 1, q \to 1$, and any sign of inhomogeneity disappears from within Hamiltonian (\ref{H2}). Hence, the classical $A = 1$ threshold holds. 
Instead, when $ \omega^{(0)} t /L \gg 1,  q \gg 1$, $ t \approx [3 \tau \left(L/ \omega^{(0)} \right)^2]^{1/3}$ evolves slowly with $\tau$. We discard all but the lowest order terms in $1/q$ in Eq. (\ref{eq:xp}) (recalling that $dq/dt$ is a mild function of time: $dq/dt \to 0, t \to 0; dq/dt \to \omega^{(0)}/L,  t \to \infty$) and arrive to the equations of motion for a pendulum with slowly varying parameters--and therefore slowly varying separatrix:
\begin{eqnarray}
d_\tau X &=& P  \nonumber  \\ 
d_\tau P &=& {A \over q^2} \cos \left[X - \omega^{(0)} t(\tau) \right]
\label{eq:xp2}
\end{eqnarray}
It is worth noting that, as $t$ and $q$ go to infinity, $X$ may grow unbounded, whereas $x$ and $\omega(x)$ stay bounded and fairly close to their initial values, see Fig. 1.  \\
Qualitatively, we may summarize the typical fate of an orbit as follows. The particle starts its trajectory as unperturbed by the wave (since $ A \ll 1$): a pure Larmor rotation. After a time $\omega^{(0)} t / L \approx O(1)$, as shown in Fig. (\ref{fig:3}), a separatrix develops in phase space of Hamiltonian. The particle crosses a first time the separatrix, remaining trapped into the corresponding potential well, then crosses it a second time, becoming passing again. These two events are genuinely non-adiabatic, hence the particle energy changes during them.

\begin{figure}[h]
\includegraphics[width=8cm]{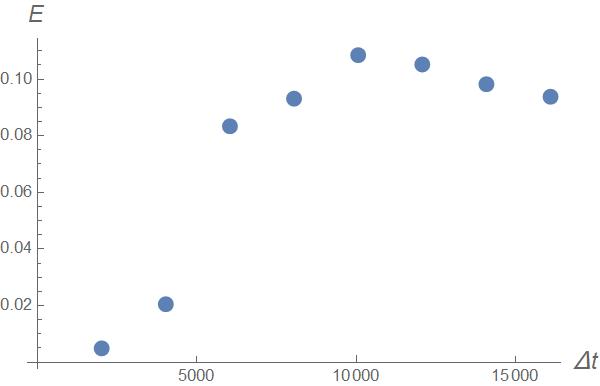}
\caption{ Mean energy $E$  gained by the particle--averaged over 600 independent initial conditions--{\it versus} the interaction time $\Delta t = t_e - t_s$, with $A = 1/5, L = 100$, and $\omega^{(0)} = 1/40$.   }
\label{fig:4}
\end{figure}

Figure (\ref{fig:4}) makes visual our conclusions: it plots the average energy gained by the particle as a function of the duration of the interaction with the wave. For the parameters employed $\omega^{(0)} t / L > 1$ when $t > 4000$ time units. Numerically, we find a transition to finite energy gains for $ t \approx 5000$. Equivalently, one could have set fixed the interaction time $\Delta t$ and varied $\omega^{(0)}$: appreciable energy gain occurs as long as $\omega^{(0)} \Delta t / L > 1$. We have checked numerically this claim, too.

We emphasize a difference with the case $\omega = $ constant, $A>1$: in this latter case there is a pulsation of the separatrix amplitude, while in the present one the separatrix moves in $p$ in phase space after emerging. Furthermore, this scenario works whatever $A$ and $L$, which proves the universality of our numerical findings. \\
By increasing the energy, one progressively pushes the trajectories farther from the origin--and therefore from the separatrix, but only orbits encompassing the phase space region swept by the separatrix are involved in the adiabatic-non adiabatic transition. Thus, if we pick initial conditions out from a thermal distribution, we get that the high-energy tail of the distribution is not affected by this mechanism: hotter particles are less heated, on average. 

\begin{figure}[h]
\includegraphics[width=8cm]{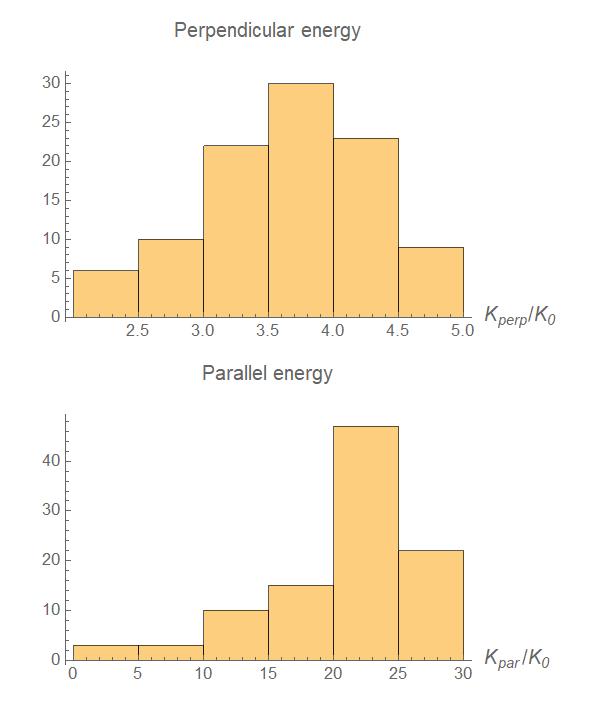}
\caption{Histograms of the final particle kinetic energy along the perpendicular direction ($K_x$) and parallel direction ($K_z$) normalized to initial kinetic energy $K_0$. 200 independent runs were performed varying randomly the initial coordinate $z$ in the interval $(0, 2 \pi)$. The other parameters used are $A = 0.1, \omega^{(0)} = 0.05, L = 25, k_z = 1, d =  24 \pi /\omega^{(0)},  t_s = 5 d, t_e = 25 d, x(0) = 0, v_x(0) = 0.01, v_z(0) = 0.01$.  }
\label{fig:5}
\end{figure}

{\it Simultaneous parallel and perpendicular heating} So far, we considered just the case of particle energization along the direction perpendicular to the magnetic field. However, there is experimental evidence that parallel heating may occur simultaneously with a perpendicular one \cite{magee11}. That paper speculated that parallel heating was a product of energy mixing along all axes by particle collisions, hence ultimately only perpendicular heating was active. However, no conclusive evidence was found, hence it is suggestive to consider the possibility that one and the same mechanism be responsible for heating both degrees of freedom. \\
Models for stochastic heating usually deal only with perpendicular degreees of freedom, although some exceptions exist (see, e.g., \cite{chandran13}). Among the several ways of introducing chaotic dynamics in the parallel direction, we considered two: 
either we added to the Hamiltonian a single plane wave travelling along ${\bf B}$ with the same frequency as the perpendicularly propagating one:  $ A_x \sin \left(x - \omega(x) t \right) +
A_z \sin \left( k_z z - \omega(x) t  \right) $, or a single wave propagating obliquely: $A \sin \left(x + k_z z - \omega(x) t \right) $ (with obvious meaning of the symbols).
The dynamics of $x$ and $z$ is thus coupled. We carried out investigations for both scenarios but report here explicit results only for the second case. Qualitative conclusions do not differ in the former case, although quantitatively (i.e., the amount of energy delivered to the particle for given conditions, and the balance  between the parallel and the perpendicular degrees of freedom) marked differences may be found. As a general rule, the larger the ratio between the parallel and perpendicular wavenumber ($k_z$ in our units), the larger the ratio between the temperature increment along the parallel and the perpendicular direction, but
investigating in detail these differences is outside the scope of the present work and will be left to further studies.  \\
Fig. (\ref{fig:5}) shows the histogram of the kinetic energy distribution along the perpendicular and parallel directions: the average final kinetic energy is about three times the initial energy along the perpendicular direction, and about twenty times along the parallel one (Recall that the {\it total} energy along the perpendicular direction is twice the kinetic energy though).  We do not repeat here the analytical study of the Hamiltonian done in the previous section, but the nature of the transition to non-adiabatic dynamics is similar. 

{\it Concluding remarks.} Summarizing, we have studied an unexplored scenario for stochastic (chaotic) particle heating, to the best of our knowledge. The conclusion we draw is that, at odds with present wisdom, heating by low-frequency waves does not necessarily require a threshold upon the wave amplitude.\\
Within the framework of Eq. (\ref{eq:1}),  inhomogeneity may affect either the wave frequency, its amplitude, the wavenumber, or several of these parameters at once. 
In the present study we investigated just the first possibility, since it is the only one which produces nontrivial consequences. Altering $\omega$ introduces a coupling between temporal and spatial coordinates. This is clearly spotted after the canonical transformations which lead to Eq. 4: in the Hamiltonian different phase space quantities appear weighted by different powers of the parameter $q(t)$. This is the reason why, eventually, the topology of the system changes, developing a separatrix.
The change of $k$ or of $A$ does not bring any such coupling.\\
Broadening a distribution function by a separatrix sweeping phase space slowly may be relevant to other inhomogeneous physical systems.  It is for instance for the motion in a Langmuir wave whose phase velocity is slowly oscillating with space.

{\it Acknowledgments.} F.S. wishes to thank B. Momo, M. Gobbin, I. Predebon, O. Pezzi, S. Cappello, and A Settino for useful discussions.

 \end{document}